\documentstyle[eqsecnum,aps]{revtex}

\begin{document}
\draft
\preprint{HEP/123-qed}
\title{resistivity peculiarities in systems with lattice distortions}
\author{Marco Zoli}
\address{Istituto Nazionale Fisica della Materia, 
Universit\'a di Camerino, 
62032 Camerino, Italy.  Fax +390577662176 -
zoli@campus.unicam.it
}

\date{Received January 15, 2001}
\maketitle
\begin{abstract}
We study a molecular lattice Hamiltonian
in which polaronic charge carriers interact with
non linear potentials
provided by local atomic fluctuations between
two equilibrium sites.
The path integral formalism is applied to select the 
class of atomic oscillations which mainly contributes 
to the partition function and the electrical resistivity 
is computed in a number of representative cases.
Non metallic resistivity behaviors are found at 
temperatures above $\simeq 100K$.


\end{abstract}


%\narrowtext
%\widetext

\setlength{\baselineskip}{20pt}   

While there is growing evidence
that fundamental properties such as the polaron size,
effective mass and ground state energy are essentially
similar in any dimension, it is still unclear 
to which extent transport properties in polaronic systems depend on the lattice structure and dimensionality. 
Besides being conceptually relevant
this question has become actual in connection with the
discovery of unusual effects in underdoped high $T_c$ superconductors.
Infact the presence of local lattice distortions with
polaron formation has been
envisaged in the high $T_c$ systems and signs of enhanced 
anharmonicity for some in- and out of plane oxygen modes
have been detected in underdoped compounds by several 
groups.
In this paper we focus on the problem of the interaction
between a polaronic quasiparticle moving through a
molecular lattice and a local structural
instability modelled by a double well potential in its
two state configuration. 
Our analysis starts from the time dependent Hamiltonian 

\begin{eqnarray}
H_{0}(\tau)=& &\, \bar \epsilon(g) \tilde c^{\dag}(\tau) 
\tilde c(\tau) + \sum_{\bf q}{\omega_{\bf q}}a^{\dag}_{\bf q}
(\tau)a_{\bf q}(\tau) + H_{TLS}(\tau) \,
\nonumber \\
& &\Bigl( H_{TLS}(\tau) \Bigr)=\, 
\left(\matrix{0 & \lambda Q(\tau) \cr 
\lambda Q(\tau) & 0 \cr} \right)\,
\nonumber \\  
& &H_{int}(\tau)=\,- 2\lambda Q(\tau) \tilde c^{\dag}(\tau) 
\tilde c(\tau) \,
\nonumber \\
& &Q(\tau)=\, -Q_o + {{2 Q_o}\over {\tau_o}}
(\tau - t_i)
\label{1}.
\end{eqnarray}

$H_0(\tau)$ is the free Hamiltonian made of: a) A polaron
created (distroyed) by $\tilde c^{\dag}(\tau) ( \tilde c(\tau))$ 
in an energy band $\bar \epsilon(g)$ whose width decreases 
exponentially by increasing the strength of the overall
electron-phonon coupling constant $g$, 
$\bar \epsilon(g)=\,Dexp(-g^2)$. b) A lattice of diatomic
molecules whose phonon frequencies
$\omega_{\bf q}$ are derived analytically for a linear chain, a square lattice and a simple cubic lattice through a
three force constants model:
the {\it intra}molecular 
frequency $\omega_0$ which largely determines the size of
the lattice distortion associated with  polaron formation,
the first ($\omega_1$) and second
($\omega_2$) neighbors {\it inter}molecular couplings
which are essential to compute the polaron properties
both in the ground state and at finite temperatures
\cite{io1}.
c) A local anharmonic potential
shaped by a Two Level System (TLS) in its 
symmetric ground state configuration.
$Q(\tau)$ is the one dimensional {\it space-time}
hopping path followed by the atom which moves between
two equilibrium positions located at $\pm Q_o$.
$\tau_o$ is the bare hopping time between the two minima 
of the TLS and
$t_i$ is the instant at which the $ith$-hop takes place.
One atomic path is characterized by the number $2n$ of hops, 
by the set of $t_i$ $(0 < i \le 2n)$ and by $\tau_o$.
The interaction is described by $H_{int}(\tau)$ with
$\lambda$ being the coupling strength between TLS and
polaron, 
$\lambda Q(\tau)$ is the renormalized tunneling 
energy.
The full partition function is obtained \cite{firenze}
after solving perturbatively the
Dyson equation for the polaron propagator and summing
over the class of linear atomic paths:

\begin{equation}
Z_T=\, Z_0\sum_{n=0}^{\infty}
\int_0^{\beta}{{dt_{2n}}\over {\tau_0}} \cdot 
\cdot
\int_0^{t_2-\tau_0}{{dt_{1}}\over {\tau_0}}
exp\Bigl[-\beta E(n,t_i, \tau_0) \Bigr]  \,
\label{2}
\end{equation}

with $E(n,t_i, \tau_0)$ being the one path atomic energy
which contains the time  retarded polaronic
interactions between successive atomic hops in the
double well potential. 
We turn now to compute the electrical resistivity $\rho(T)$
due to the polaronic charge carrier scattering by
the impurity potential 
provided by the TLS. The input parameters of the model are 
the three molecular force constants, $g$, $D$ and the bare
energy $\lambda Q_0$.
$Q_0$ can be chosen as $\simeq 0.05 \AA$ consistently with
reported values in the literature on TLS's. The bare electronic 
band $D$ is fixed at 0.1eV.
In Fig.1, we take a 3D lattice with rather large phonon 
frequencies but still 
in moderately adiabatic conditions, an intermediate 
e-ph coupling which ensures polaron mobility and tune 
the TLS-polaron coupling $\lambda$. 
A resistivity peak located at $T \simeq 150K$ arises
at $\lambda \ge 700meV \AA^{-1}$ with height and 
width of the peak being strongly dependent on $\lambda$
hence, on the TLS energy. The low $T$ resistivity
displays the maximum at the unitary limit thus revealing
that TLS's are at work here while
the high $T$ ($T > 300K$) behavior can be metallic like (
$\lambda < 800meV \AA^{-1}$) or semiconducting like 
($\lambda > 800meV \AA^{-1}$). 
$\lambda \simeq 700 - 800meV \AA^{-1}$ corresponds to a
TLS energy of $\simeq 35 - 40meV$ which is comparable
to the value of the bare polaron energy
band $\bar \epsilon(g=1) \simeq 37meV$. 
While the peak does not shift substantially by varying 
the strength of the intermolecular forces its height 
turns out to be rather sensitive to those strength 
hence, to the size of the quasiparticle.
At larger $\lambda$'s the
resonance peak is higher and broader since an 
increasing number of incoming
polarons can be off diagonally scattered by the TLS.
However, the appearance of this many body effect 
mediated by
the local potential does not change the position 
of the peak either, which
instead can be shifted towards lower temperatures
(Fig.2) by reducing substantially $\omega_0$. 
The non metallic behavior at $T$ larger than $\simeq 90K$
reminds of the anomalous c-axis resistivity observed in underdoped
high $T_c$ superconductors \cite{ito}.
Anharmonic features of the oxygen modes have been recognized to be 
larger in underdoped samples and doping dependent 
polaron
formation has been correlated to
distortions of the oxygen environment. 
We suggest that the semiconducting like $\rho_c$
in underdoped high $T_c$ superconductors might be ascribed to 
anharmonic
potentials due to oxygen displacements strongly coupled via 
$\lambda$
(as in Fig.2) to polaronic 
carriers. To attempt a comparison with experiments 
we need to fix the residual resistivity. By extrapolating to 
$T=\,0$ the normal state data on $YBa_2Cu_3O_{7-\delta}$ one 
derives $\rho(T=\,0) \simeq 3m\Omega cm$, hence the experimental 
peak value $\rho \simeq 20m\Omega cm$ observed in the 80K compound $YBa_2Cu_3O_{6.87}$ \cite{ito} can be reproduced in my model by
$\lambda \simeq 1300meV \AA^{-1}$ which corresponds to a local 
mode energy of $\simeq 65meV$, in fair agreement with the measured 
energies of phonons strongly coupled to the charge carriers.
In the range $80K < T < 200K$ the 
$\lambda \simeq 1300meV \AA^{-1}$ curve
fits rather well the data of ref. \cite{ito}. At $T=\,200K$, 
the experimental $\rho$ is $\simeq 10m\Omega cm$ 
and the calculated 
value is $\simeq 8m\Omega cm$. In the range 
$200K < T < 300K$, the experimental $\rho$ is essentially flat
while my results still exhibit a $d\rho/dT < 0$ behavior.

In conclusion,
strength of the overall electron-phonon coupling ($g$),
strength of the polaron-local potential coupling ($\lambda$)
and strength of the molecular forces interfere, giving rise to
a variety of resistivity behaviors versus temperature.
As a main feature, when the conditions for resonant scattering 
between polaron and
local double well potential are fulfilled, a broad resistivity peak
shows up in the 100-150K range. 
However, the shape of the peak essentially
depends on the dynamics of the local potential 
and, for sufficiently low atomic tunneling energies, metallic
conductivity conditions are recovered.
Generally, in 3D the resistivity maximum is less pronounced than
in low dimensionality and, for sufficiently strong intermolecular
couplings, the height of the peak lies below the residual
resistivity value, being therefore hardly visible in experiments.

\pagebreak

\section*{figure captions}

\begin{figure}
\vspace*{5.5truecm}
\caption{Electrical Resistivity normalized to the residual
resistivity for five values of the polaron-TLS coupling
$\lambda$. $g=\,1$. The force constants which control the phonon
spectrum are: $\omega_0=\,100meV, \omega_1=\,20meV,
\omega_2=\,10meV$}
\end{figure}

\begin{figure}
\vspace*{5.5truecm}
\caption{Electrical Resistivity normalized to the residual
resisitivity  for nine values of the polaron-TLS coupling
$\lambda$. $g=\,1$.  $\omega_0=\,50meV, \omega_1=\,20meV,
\omega_2=\,10meV$. The experimental data (X) are taken 
from ref.[3]}
\end{figure}

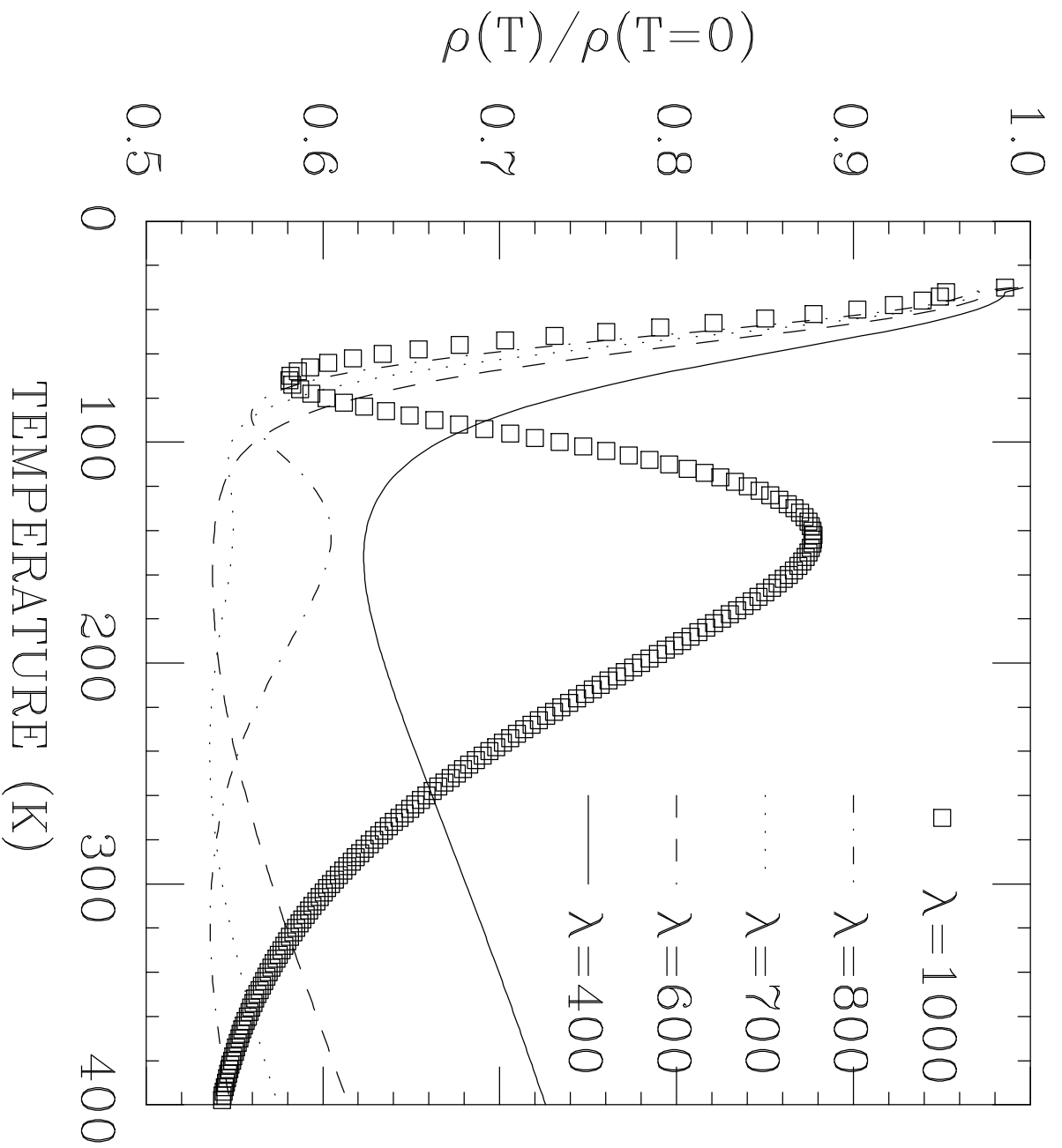
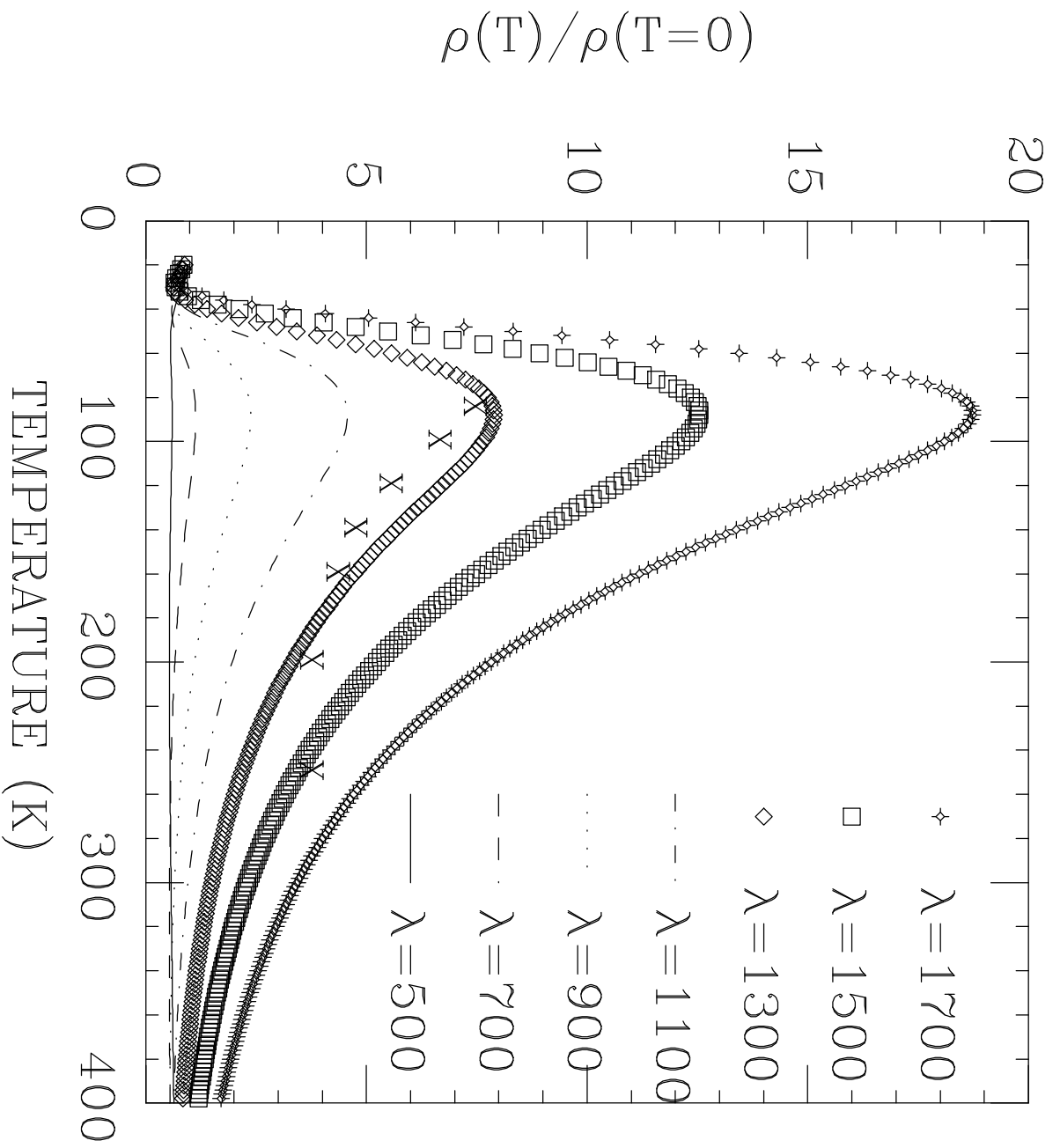
\end{document}